\documentclass[aps,prl,twocolumn,showpacs,superscriptaddress,groupedaddress]{revtex4-1} 
\usepackage{dcolumn}
\usepackage{bm}   
\usepackage{xcolor}
\usepackage[utf8]{inputenc}
\usepackage{amsmath, amssymb}
\usepackage{graphicx}
\usepackage[english]{babel}

\begin{document}

\title{Neutrino Decay and Solar Neutrino Seasonal Effect}
\author{R. Picoreti}\affiliation{Instituto de F\'isica Gleb Wataghin - UNICAMP, {13083-859}, Campinas SP, Brazil}
\author{M. M. Guzzo}\affiliation{Instituto de F\'isica Gleb Wataghin - UNICAMP, {13083-859}, Campinas SP, Brazil}
\author{P. C. de Holanda}\affiliation{Instituto de F\'isica Gleb Wataghin - UNICAMP, {13083-859}, Campinas SP, Brazil}
\author{O. L. G. Peres} \affiliation{Instituto de F\'isica Gleb Wataghin - UNICAMP, {13083-859}, Campinas SP, Brazil} \affiliation{The Abdus Salam International Centre for Physics, Trieste, Italy}

\begin{abstract}
We consider the possibility of solar neutrino decay as a sub-leading effect on their propagation between production and detection. Using current oscillation data, we set a new lower bound to the $\nu_2$ neutrino lifetime at $\tau_2\, /\, m_2 \geq 7.2 \times 10^{-4}\,\,\hbox{s}\,.\,\hbox{eV}^{-1}$ at $99\%\,$C.L.. Also, we show how seasonal variations in the solar neutrino data can give interesting additional information about neutrino lifetime.
\end{abstract}
\maketitle

\section{Introduction}

Beyond any reasonable doubt, it is now established that neutrinos have non-zero, non-degenerate masses and, thus, it would be possible - if not mandatory - for them to decay into other particles. 

Before the establishment of the LMA-MSW solution~\cite{GonzalezGarcia:2002dz}, decay was studied both by itself and in combination with standard flavor oscillations to explain the difference between the expected solar neutrino flux from nuclear fusion processes in the Sun and the detected flux on Earth - the so-called Solar Neutrino Problem (SNP).

Although it is now ruled out as a leading process~\cite{Bandyopadhyay:2002qg}, one can investigate neutrino decay as a sub-leading effect in the propagation of solar neutrinos and set limits to their lifetime using the most recent experimental data.   

Solar neutrinos are produced in the nuclear fusion processes that power the Sun. In such processes, Hydrogen nuclei are converted into Helium through several intermediate reactions, among which some produce neutrinos in very particular spectra - both continuous and monochromatic.

Over the years, several experiments were developed for the detection of solar neutrinos at different energy ranges. From the pioneer Homestake~\cite{Cleveland:1998nv} chlorine experiment - which first hinted at the SNP - through the gallium experiments GALLEX~\cite{Hampel:1998xg}, SAGE~\cite{Abdurashitov:2009tn} and GNO~\cite{Altmann:2005ix} to the water Cherenkov detectors Kamiokande, Super Kamiokande~\cite{Abe:2010hy} and SNO~\cite{Aharmim:2011vm}. Most recently, the Borexino~\cite{Bellini:2013lnn} experiment also measured the $^7$Be line.

The LMA-MSW solution, now considered the best explanation for the SNP, in combination with the measurement of the other oscillation parameters by experiments designed for atmospheric, reactor and long-baseline neutrinos established the scenario of three massive light neutrinos that mix~\cite{GonzalezGarcia:2002dz}. With such precise measurements of the standard oscillation parameters, it is possible to investigate new phenomena such as the neutrino decay scenario: $\nu^{\prime}\rightarrow\nu+X$.

For solar neutrinos, the decay of the mass eigenstate $\nu_2$ into the lighter state $\nu_1$ is disfavored by the data and the current bound to $\nu_2$ lifetime for invisible non-radiative decays~\cite{Bandyopadhyay:2002qg} is $\tau_2/m_2 \geq 8.7 \times 10^{-5}\,\,\hbox{s}\,.\,\hbox{eV}^{-1}$ at $99\%$ C.L.. 
Similarly, from the combined accelerator and atmospheric neutrino data, $\nu_3$ lifetime is $\tau_3/m_3 \geq 2.9 \times 10^{-10}\,\,\hbox{s}\,.\,\hbox{eV}^{-1}$ at $90\%$ C.L.~\cite{GonzalezGarcia:2008ru}.
Also, an analysis of long-baseline experiments MINOS and T2K give a combined limit of $\tau_3/m_3 \geq 2.8\times 10^{-12}\,\,\hbox{s}\,.\,\hbox{eV}^{-1}$ at $90\%$ C.L.~\cite{Gomes:2014yua}.

In this work, we consider the decay scenario in which all the final products are invisible. We combine the available solar neutrino data with KamLAND~\cite{Gando:2010aa} and Daya Bay~\cite{An:2013zwz} data. For both experiments, the effect of neutrino decay is minimum thus allowing us to constrain the standard neutrino mixing parameters independently of the decay parameter $\tau_2/m_2$ and leading us to obtain a robust bound on $\nu_2$ lifetime. Additionally, we show how seasonal variations in the solar neutrino data, which are enhanced by neutrino decay, can give some interesting information about neutrino lifetime.

\section{Formalism}

After production in the solar core, neutrinos propagate outwards undergoing flavor oscillation and resonant flavor transition due to the solar matter potential. After emerging from the solar matter, they travel across the interplanetary medium until they reach the Earth's surface where they can be detected promptly (during the day) or after traversing Earth's matter (during the night - on which they may also be subject to matter effects).   

The transition amplitude for an electron neutrino produced in the Sun to be detected on Earth as a neutrino of flavor $\alpha$, $\nu_e\to \nu_{\alpha}$, for the standard case of neutrino oscillations with MSW effect, can be written as~\cite{Joshipura:2002fb}: 
\begin{equation}
A_{e\alpha} = \sum A^{\odot}_{e i}\, A_{ii}^{\rm vac}\, A^{\oplus}_{i \alpha}\,,
\end{equation}
where $A^{\odot}_{ei}$ is the transition amplitude of an electron neutrino produced in the solar core to be in a $\nu_i$ state in the solar surface, $A_{ii}^{\rm vac}$ is the propagation amplitude between Sun and Earth surfaces, and $A^{\oplus}_{i\alpha}$ is the transition amplitude of a $\nu_i$ to be in a $\nu_e$ state upon detection on Earth. 

Considering the current limits to their lifetime, 
neutrinos do not decay inside the Sun. It suffices to consider their decay on the way to Earth by taking the decay survival probability for invisible decay, of a neutrino mass-eigenstate $i$, with energy $E_\nu$, after propagating a distance $L$, to be: 
\begin{equation}
P_{\rm i}^{\rm dec} = \exp\left[-\left(\frac{\alpha_i}{E_\nu}\right)\,L\right]\,,\,\, \hbox{with}\,\, \alpha_i = \frac{m_i}{\tau_i}\,,
\label{decay}
\end{equation}
where $m_i$ is the eigenstate's mass , $\tau_i$ is the eigenstate's lifetime and $L$ is the Sun-Earth distance.

For the assumption that only the $\nu_2$ mass-eigenstate is unstable, the electron neutrino survival probability including decay and oscillation is: 
\begin{equation}
 P(\nu_e\to \nu_e)= c^4_{13}\, \Big[P^{\odot}_{e 1}\, P^{\oplus}_{1e}\,+\,P^{\odot}_{e 2}\,\left( P_{2}^{\rm dec}\right) \, P^{\oplus}_{2e}\Big] + s^4_{13}\,,
\label{eq-surv}
\end{equation}
where $s_{ij} = \sin\theta_{ij}$ and $c_{ij} = \cos\theta_{ij}$ and $P_{\rm i}^{\rm dec}$ is given in Eq.~(\ref{decay}). In this scenario, one interesting point is that the sum over all probabilities is not equal to $1$, as explicitly we have: 
\begin{equation}
\sum_{\alpha=e,\mu,\tau} P(\nu_e\to \nu_{\alpha})= 1 - c_{13}^2\,P_{e2}^\odot\,\left(1-P_2^{\rm dec}\right)\,.
\label{eq-prob}
\end{equation}
This non-unitary evolution was discussed in Ref.~\cite{Berryman:2014yoa}. For this scenario, we should compute 
$P(\nu_e\to \nu_e)$ and $\sum_{\alpha=\mu,\tau} P(\nu_e\to \nu_{\alpha})$ that are independent probabilities.

One important point is that, for appreciable values of $\tau_2/m_2$, the solar neutrino data can be explained by a combination of standard three neutrino MSW oscillation and decay, which leads to a degenerescence between neutrino parameters, specially $\Delta m_{21}^2$ and $\tau_2/m_2$~\cite{Bandyopadhyay:2002qg} .

\section{Analysis and Results}

For the analysis of $\nu_2$ decay over the Earth-Sun distance and how it affects the expected rate for each solar neutrino experiment, we calculate the neutrino survival probabilities as shown in Eq.~(\ref{eq-surv}) and Eq.~(\ref{eq-prob}), numerically, under the assumption of adiabatic evolution inside the Sun~\cite{deHolanda:2010am}. Then, 
we compute the expected event rate for each relevant experiment and compare it to their data.

We include Homestake total rate~\cite{Cleveland:1998nv}, GALLEX and GNO combined total rate~\cite{Kaether:2010ag}, SAGE total rate~\cite{Abdurashitov:2009tn}, SuperKamiokande I full energy and zenith spectrum~\cite{Hosaka:2005um},
SNO combined analysis~\cite{Aharmim:2011vm} and Borexino 192-day low-energy data~\cite{Arpesella:2008mt}. Then, we build a $\chi^2$ function as a function of the relevant parameters ${\chi^2_{\odot}=\chi^2_{\odot}(\tan^2\theta_{12}, \Delta m^2_{21}, \sin^2\theta_{13}, \tau_2/m_2)}$. 

We can add complementary information from the reactor experiments KamLAND~\cite{Gando:2010aa} and Daya Bay~\cite{An:2013zwz} and their detection of $\bar{\nu}_e$ oscillations. {\em One important point that led us toward this analysis} is the fact that these reactor experiments give precise constraints on $\Delta m^2_{21}$ and $\sin^2\theta_{13}$. KamLAND and Daya Bay have typical baselines of ${L/E_{\nu}\sim 10^{-10}}\,\,\hbox{s}\,.\,\hbox{eV}^{-1}$ and ${\sim 10^{-12}}\,\,\hbox{s}\,.\,\hbox{eV}^{-1}$ respectively. For the currently allowed values of $\tau_2/m_2$, one has that $P^{\rm dec}_i\sim 1$, which implies that, in the context of these experiments, decay can be neglected and the relevant neutrino probability is the standard three neutrino expression $P(\bar{\nu}_e\rightarrow\bar{\nu}_e) = P_{\bar{e}\bar{e}}$, with: 
\begin{eqnarray}
P_{\bar{e}\bar{e}} = 1 -c^4_{13}S^{2}_{12}\sin^2\Delta_{21} -\, S^{2}_{13}\sin^2{\Delta m^2_{ee}}\,,
\end{eqnarray}
where $S_{ij} = \sin 2\theta_{ij}$, $\Delta_{ij}=\delta m_{ij}^2/4 E_{\nu}$ and $\delta m_{ij}^2\equiv m^2_i-m^2_j$, and we define an effective mass square difference $\sin^2\Delta m^2_{ee}\equiv c^2_{12}\sin^2\Delta_{31} + s^2_{12}\sin^2{\Delta_{32}}$.

\begin{figure}[t!]
	\centering
		\includegraphics[width=.4\textwidth]{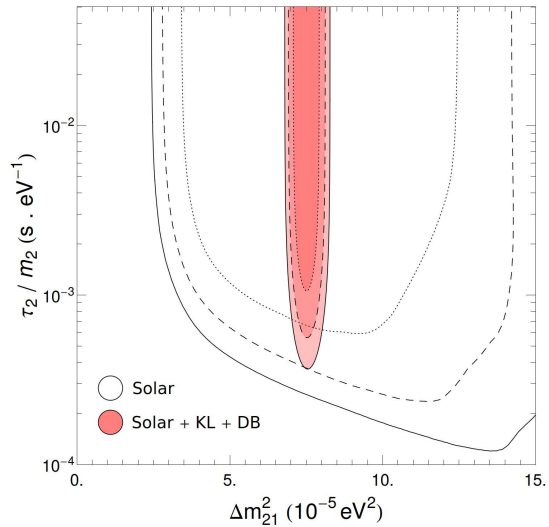}
	 \caption{Allowed regions for the decay parameter $\tau_2/m_2$ and the mass squared difference $\Delta m^2_{21}$. The hollow curves represent the solar neutrino analysis only and the filled curves represent the combined analysis of solar, KamLAND and Daya Bay data. The dotted, dashed and continuous line represent respectively $90\%$ C.L., $99\%$ C.L. and $99.9\%$ C.L..} 
\label{fig1}
\end{figure}

This implies that the standard neutrino analysis for three neutrinos of KamLAND and Daya Bay experiments can also be used for decay scenario. In other words, we can identify $\chi^2_{\rm decay} = \chi^2_{\rm no \,decay}$ in our analysis for both experiments. The KamLAND experiment have provided a $\chi^2_{\rm KL}$ function for the standard three neutrino scenario used in Ref.~\cite{Gando:2010aa} and available in table format as a function of $\tan^2\theta_{12}$, $\Delta m^2_{21}$ and $\sin^2\theta_{13}$. For the Daya Bay experiment, the $\chi^2_{\rm DB}$ function is available in table format provided in the supplementary material from Ref.~\cite{An:2013zwz} as a function of $\Delta m^2_{ee}$ and $\sin^2\theta_{13}$. 

Then, we write the combined $\chi^2$ function for solar, KamLAND and Daya Bay data as: 
\begin{eqnarray}
\chi^2 & = & \chi^2_{\odot}(\tan^2\theta_{12}, \Delta m^2_{21}, \sin^2\theta_{13}, \tau_2/m_2) + \nonumber \\ 
      & + & \chi^2_{\rm KL}(\tan^2\theta_{12}, \Delta m^2_{21}, \sin^2\theta_{13}) + \nonumber \\
       & + & \chi^2 _{\rm DB}(\Delta m^2_{ee}, \sin^2\theta_{13})\,, 
       \label{chi2}
\end{eqnarray}
where $\Delta m^2_{ee}$ was defined before and over which we can promptly marginalize the $\chi^2$. From Eq.~(\ref{chi2}), we find the allowed regions for independent parameters $\tan^2\theta_{12}$, $\sin^2\theta_{13}$, $\Delta m^2_{21}$, and $\tau_2/m_2$. By marginalizing over the first two, we get the allowed region for the mass squared difference $\Delta m^2_{21}$ and the decay parameter $\tau_2/m_2$ as shown in Fig.~\ref{fig1}, where the hollow (filled) regions show the results for the solar neutrino (combined) analysis. 

\begin{figure}[t!]
	\centering
		\includegraphics[width=.4\textwidth]{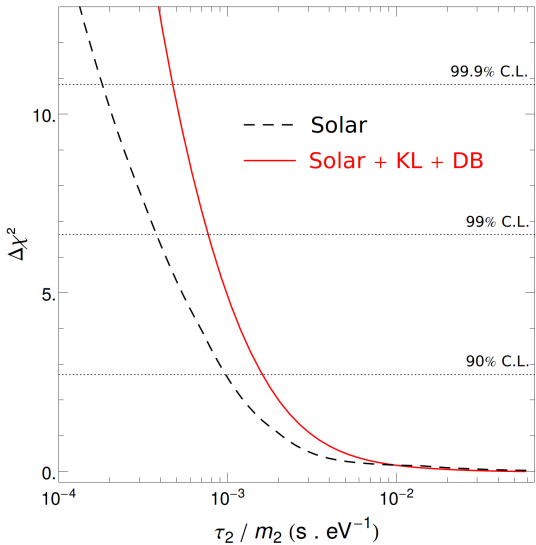}
		 \caption{$\Delta \chi^2$ for $\nu_2$ lifetime $\tau_2/m_2$. The black (red) curve shows only solar (combined) data analysis.} 
\label{fig2}
\end{figure}

The degenerescence between $\Delta m^2_{21}$ and $\tau_2/m_2$ is evident in the hollow regions of Fig.~\ref{fig1}, where higher (lower) values of $\Delta m^2_{21}$ and lower (higher) values of $\tau_2/m_2$ are allowed. High values of $\Delta m^2_{21}$ are ruled out in the standard neutrino scenario because it leads to spectral distortions that are disfavored by the solar neutrino data. On the other hand, it could be turned into a viable solution at the cost of having lower values of $\tau_2/m_2$. The inclusion of KamLAND and Daya Bay data break this degenerescence due to their precise independent measurement of $\Delta m_{21}^2$ and $\sin^2 \theta_{13}$ respectively.
We can now precisely isolate the contribution of the decay parameter $\tau_2/m_2$. The complete marginalization over the standard parameters results in the curve shown in Fig.~\ref{fig2} of $\Delta \chi^2$ as a function of $\tau_2/m_2$. From it, we can extract a lower limit to the $\nu_2$ eigenstate lifetime: 
\begin{equation}
\tau_2\, /\, m_2 \geq 7.7 \times 10^{-4}\,\,\hbox{s}\,.\,\hbox{eV}^{-1},\,\hbox{at 99\% C.L.}\,,
\end{equation}
which corresponds to the upper bound to the decay parameter $\alpha_2 \leq 8.5 \times 10^{-13}\,\hbox{eV}^{2}$. 
\begin{table}[t!]
\begin{tabular}{ccc}\vspace{.1cm}
Experiment & $\epsilon_{\rm exp} \pm \sigma_{exp}$ & $\,\left(\epsilon_{\rm exp} \pm \sigma_{exp}\right)/\epsilon_0\,$ \\ \hline 
Borexino~\cite{Bellini:2013lnn} & $0.0398 \pm 0.0102$ & $2.38 \pm 0.61$  \\ \hline
SuperKamiokande-I ~\cite{Smy:2003jf} & $0.0252 \pm 0.0072$ & $1.51 \pm 0.43$  \\ \hline 
SNO Phase I ~\cite{Aharmim:2005iu} & $0.0143 \pm 0.0086$ & $0.86 \pm 0.51$ \\ \hline
\end{tabular}
\caption{Experimental best-fit values and errors for Earth's orbital eccentricity $\epsilon$ for different solar neutrino experiments. We also show the ratio between the fitted values and the Earth's eccentricity $\epsilon_0$.}
\label{tab1}
\end{table}
\section{Seasonal Effect}
One interesting consequence of the decay scenario that has never been explored is its effect in the seasonal variation of solar neutrino flux. 

In the absence of decay, the neutrino flux arriving on Earth is given by $\phi_\nu^\oplus = \phi_\nu^\odot / (4\pi r^2)$, where $r = r(t)$ is the time-dependent Earth-Sun distance. The ratio between maximum (perihelion) and minimum (aphelion) fluxes is ${R_0 = (1+\epsilon_0)^2/(1-\epsilon_0)^2}$, where $\epsilon_0=0.0167$ is the eccentricity of Earth's orbit.

The inclusion of decay modifies the ratio between maximum and minimum neutrino fluxes and hence also the expected eccentricity $\epsilon$ as given by:
\begin{equation}
R = R_0\,\frac{N(r_{\rm min})}{N(r_{\rm max})} = \frac{(1+\epsilon)^2}{(1-\epsilon)^2}\,,
\label{epsilon}
\end{equation}
where $r_{\rm max}(r_{\rm min})$ is the aphelion (perihelion) distance and $N$ is the number of events calculated from the convolution of the adequate probabilities and cross sections for each experiment. 

From Eq.~(\ref{eq-surv}) and Eq.~(\ref{eq-prob}), we know that $N(r_{\rm min})~>~N(r_{\rm max})$ due to $P_2^{\rm dec}$ dependence on the orbital distance. This implies that $R>R_0$ for any neutrino energy and thus, for any neutrino decay scenario, an enhancement in the seasonal variation of the solar neutrino flux would be expected.

Thus, the measurement of an eccentricity $\epsilon > \epsilon_0$ is a hint in the direction of the neutrino decay scenario. In fact, some experiments have measured Earth's orbital eccentricity to be different than the standard value albeit still compatible with $\epsilon_0$ as shown in Table~(\ref{tab1}). 

\begin{figure}[t!]
\begin{minipage}[c]{.1\linewidth}
\centering{\includegraphics[width=2.2\textwidth]{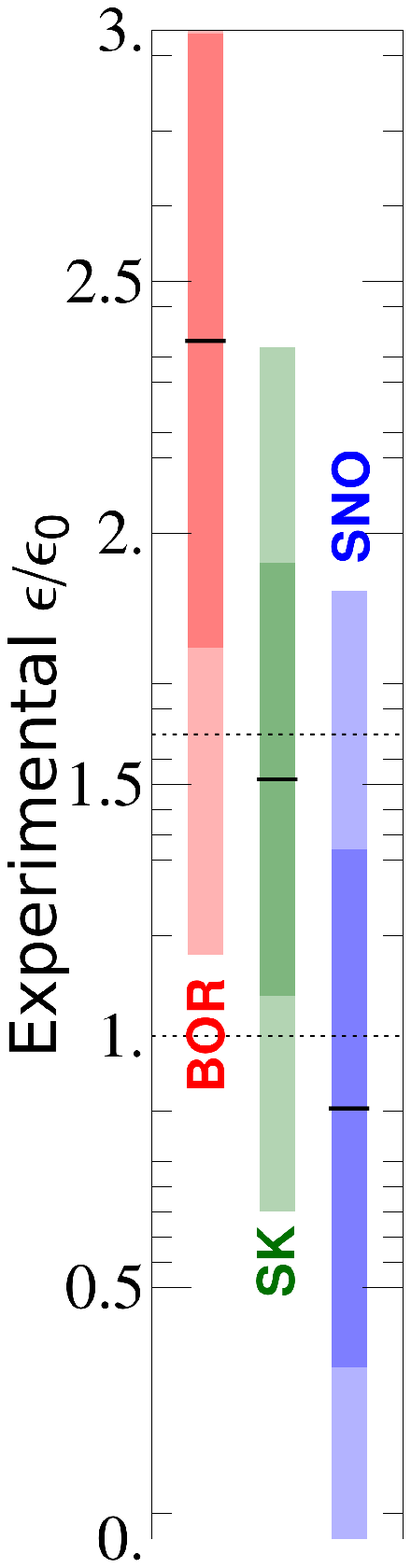}}
\end{minipage}
\hspace{.6 cm}
\begin{minipage}[c]{.8\linewidth}
\centering{\includegraphics[width=.9\textwidth]{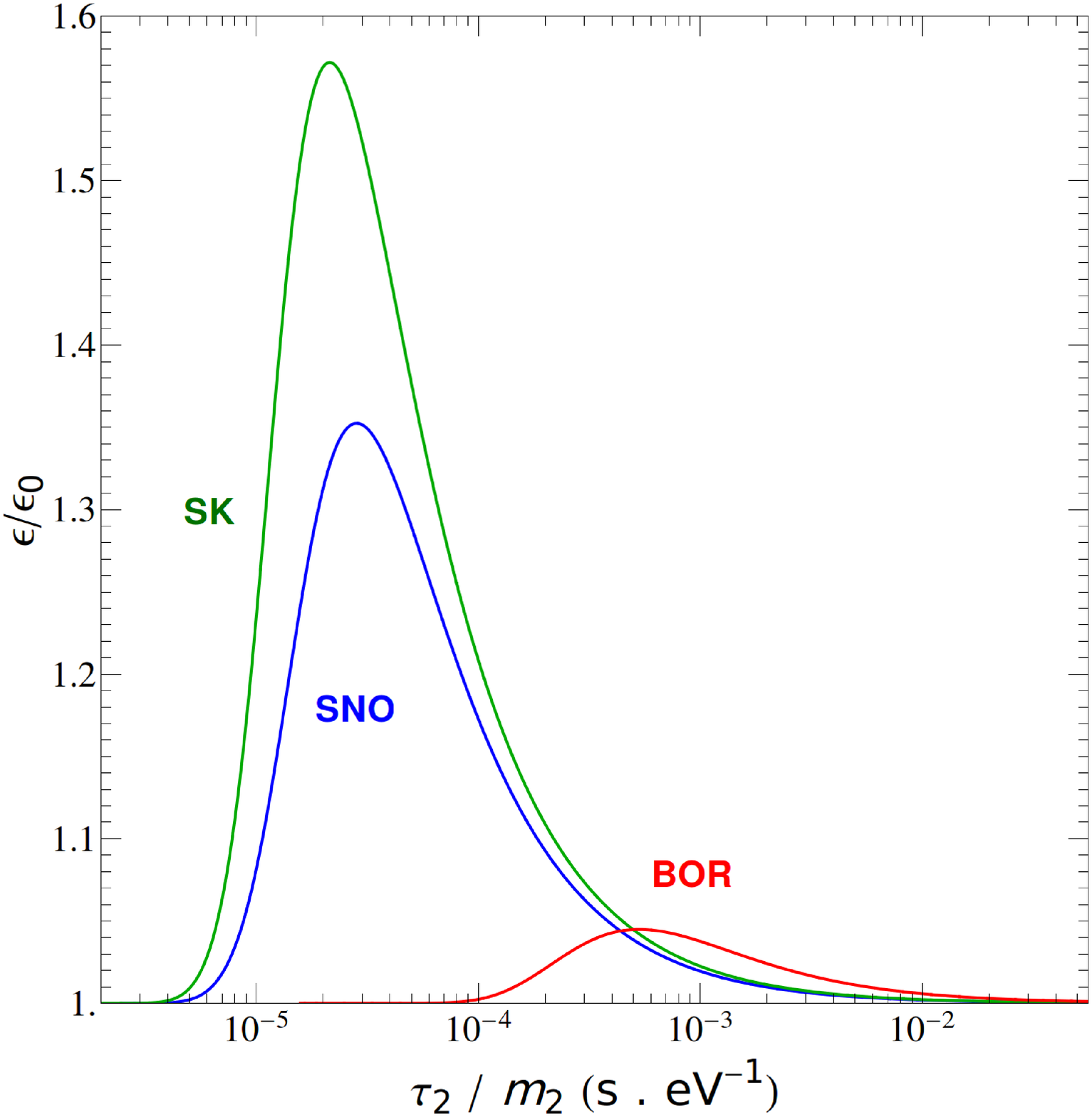}}
\end{minipage}
	 \caption{Left: Experimental values for $\epsilon/\epsilon_0$. Black lines are the best-fit values and darker (lighter) shades are the $1\sigma$ ($2\sigma$) ranges as shown in Table~(\ref{tab1}). Right: Dependence of the orbital eccentricity $\epsilon$ with the neutrino lifetime $\tau_2/m_2$ as it would be measured by different experiments - the $^7$Be line in Borexino (BOR) in red, Super-Kamiokande (SK) in green, and SNO in blue.} 
	 \label{fig3}
\end{figure}

Fig.~\ref{fig3} shows the dependence of the neutrino eccentricity $\epsilon$ with the neutrino lifetime $\tau_2/m_2$ as it would be measured by different experiments.
As it can be seen, the lower (higher) the energy of the neutrinos, the smaller (bigger) is the enhancement in the seasonal variation. Since the $\nu_2$ content in the neutrino flux leaving the Sun depends on the energy, one has that for lower (higher) energies, there are less (more) $\nu_2$ available to decay and hence the seasonal variation will be smaller (bigger). Also, due to the decay survival probability, the lower (higher) the energy, the bigger (smaller) is the lifetime at which the enhancement is maximum.

We can include the eccentricity data in the analysis with a penalty function added to the $\chi^2$ for each experiment: $\chi^2_{\rm seasonal}=(\epsilon_{\rm exp}-\epsilon)^2/(\sigma_{\rm exp})^2$. It results in a slightly lower value:
\begin{equation}
\tau_2\, /\, m_2\geq 7.2 \times 10^{-4}\,\,\hbox{s}\,.\,\hbox{eV}^{-1},\,\hbox{at 99\% C.L.}\,
\end{equation}
due to the fact that the current eccentricity measurements and errors will favor lower, already excluded, lifetimes, for which the enhancement in the seasonal variation (and hence measured eccentricity) is higher.

\section{Conclusion}
We know that neutrinos oscillate with non-zero masses and mixing angles. Can neutrinos decay? The answer is negative from the combined analysis of data of solar neutrino experiments and KamLAND and Daya Bay data. From our analysis, we have obtained a new upper bound to the $\nu_2$ eigenstate lifetime $\tau_2\, /\, m_2 \geq 7.2 \times 10^{-4}\,\,\hbox{s}\,.\,\hbox{eV}^{-1}$ at $99\%$ C.L.. which is almost one order higher than the previous bound~\cite{Bandyopadhyay:2002qg} at $\tau_2/m_2 \geq 8.7 \times 10^{-5}\,\,\hbox{s}\,.\,\hbox{eV}^{-1}$ at $99\%$ C.L.. 

Also, we have shown how decay can enhance the seasonal variation of solar neutrino fluxes and how it affects the measurement of Earth's orbital eccentricity. Current data is not good enough to improve the constraints to neutrino lifetime. Although future experiments could certainly improve on the measurement of solar neutrino fluxes and thus better constrain neutrino lifetime, the analysis of existing data from later phases of, e.g., Super-Kamiokande and SNO for its seasonal variation could, in principle, already improve such constraints. We urge those experimental collaborations~\cite{Smy:2003jf, Aharmim:2005iu} to redo their analysis with more of the available data.

\textbf{Added Note:} During the final stages of this work, a paper~\cite{Berryman:2014qha} was released which argues for $\alpha_2 < 9.3 \times 10^{-13}\,\,\hbox{eV}^{2}$ at $2\sigma$ (which corresponds to $\tau_2\, /\, m_2 \geq 7.1 \times 10^{-4}\,\,\hbox{s}\,.\,\hbox{eV}^{-1}$). Our result at $2\sigma$ is $\alpha_2 < 5.5 \times 10^{-13}\,\,\hbox{eV}^{2}$  (which corresponds to $\tau_2\, /\, m_2 \geq 1.2 \times 10^{-3}\,\,\hbox{s}\,.\,\hbox{eV}^{-1}$) which is similar to, but a more constrained neutrino lifetime than Ref.~\cite{Berryman:2014qha}.

\section{Acknowledgments}
The authors would like to thank FAPESP, CNPq and CAPES for several financial supports. O.L.G.P. thanks the support of FAPESP funding grant 2012/16389-1.

\bibliography{nudec.bib}

\end{document}